\documentstyle[preprint,prd,aps]{revtex}
\global\arraycolsep=2pt
\makeatletter
\def\fmslash{\@ifnextchar[{\fmsl@sh}{\fmsl@sh[0mu]}}
\def\fmsl@sh[#1]#2{%
  \mathchoice
    {\@fmsl@sh\displaystyle{#1}{#2}}%
    {\@fmsl@sh\textstyle{#1}{#2}}%
    {\@fmsl@sh\scriptstyle{#1}{#2}}%
    {\@fmsl@sh\scriptscriptstyle{#1}{#2}}}
\def\@fmsl@sh#1#2#3{\m@th\ooalign{$\hfil#1\mkern#2/\hfil$\crcr$#1#3$}}
\makeatother

\begin{document}
\draft\pagenumbering{roma}
%%%%%%%%%%%%%%%%%%%%%%%%%%%%%%%%%%%%%%%%%%%%%%%%%%%%%%%%%%%%%%%%%%%%%%%%%%%%%
\title{ QCD sum rule analysis of the subleading Isgur-Wise form-factor $\tau_1(v\cdot v')$ and
$\tau_2(v\cdot v')$ for $B\to D_1\ell\bar\nu$ and $B\to D_2^*\ell\bar\nu$  }
\author{ Ming-Qiu Huang, Cheng-Zu Li} 
\address{CCAST (World Laboratory) P.O. Box 8730, Beijing, 100080}
\address{and Department of Applied Physics, Changsha Institute of Technology,
Hunan 410073, China}
\author{Yuan-Ben Dai}
\address{Institute of Theoretical Physics, Academia Sinica,
P.O.Box 2735, Beijing 100080, China}
\date{ \today}
\maketitle
\thispagestyle{empty}
\vspace{15mm}
\begin{abstract}
We present a QCD sum rule calculation of the  Isgur-Wise form-factor $\tau_1(v\cdot v')$ and
$\tau_2(v\cdot v')$ for the semileptonic decays $B\to D_1(2420)\ell\bar\nu$ and 
$B\to D_2^*(2460)\ell\bar\nu$ in the framework of heavy quark
effective theory. These two universal functions, associated with the  matching of
the weak currents in QCD onto those in the effective theory, appear at the order $1/m_Q$ in the heavy quark 
expansion of meson weak decay form factors.

\end{abstract}
\vspace{4mm}
\pacs{PACS number(s): 14.40.-n, 12.39.Hg, 11.55.Hx, 12.38.Lg}
 
\vspace{3.cm}
%\noindent
%April 1998
\newpage
\pagenumbering{arabic}
%%%%%%%%%%%%%%%%%%%%%%%%%%%%%%%%%%%%%%%%%%%%%%%%%%%%%%%%%%%%%%%%%%%%%%%%%%%%%%%%
\section{Introduction}
\label{sec1} 

The heavy quark effective theory (HQET) \cite{HQET,neubert1} has important 
consequences on  the spectroscopy and the  decay matrix 
elements of the hadrons containing a single heavy quark $Q$.
In the infinite mass limit, the spin and parity of the heavy quark and that of the 
light degrees of freedom are separately conserved. This allows that the hadronic states
can be classified  in degenerate doublets by the total angular momentum $j$ and the 
angular momentum of the light degrees of freedom $j_\ell$. In the case of $\bar q Q$ mesons,
coupling  $j_\ell$ with the spin of heavy quark $s_Q=1/2$ yields  a doublet  with total 
spin $j=j_\ell\pm 1/2$. The ground state mesons with 
$j_\ell^{P}=\frac12^-$ are the doublet ($D$,$D^*$) for $Q=c$ and ($B$,$B^*$) for $Q=b$. 
The excited heavy mesons with $j_\ell^P=1^+/2$ and $3^+/2$ can be classified 
in two  doublets of spin symmetry ($0^+$,$1^+$) and ($1^+$,$2^+$), respectively.
The charmed mesons with $j_\ell^P=3^+/2$ have been 
experimentally observed. For $q=u$,$d$, these mesons are denoted as the 
$D_1(2420)$ and the $D_2^*(2460)$ mesons. While the $j_\ell^P=1^+/2$ doublet ($D'_0$, $D'_1$) 
has not been observed yet. 

One of the most important applications of the heavy quark symmetries has
been the study of semileptonic transitions between two heavy hadrons.
The hadronic matrix elements of weak currents between members of the doublets
identified by $j_\ell$ and $j_{\ell'}$ can be expressed in terms of universal
form factors which are funtions of the dot-product, $y=v\cdot v'$, of the
initial and final hadron four-velocities. A well-known result 
is that the semileptonic decays $B\to Dl\bar\nu$ and
$B\to D^* l\bar\nu$, in the $m_Q\to\infty$ limit,can be described in terms of a single
universal function, the Isgur-Wise function $\xi(y)$. 
 In the case of transitions between members belonging
to different heavy quark multiplets, additional form factors needs to be introduced.
For example, for the $B$ semileptonic decay into excited charmed mesons
($D'_0$, $D'_1$) and ($D_1$, $D^*_2$) the weak matrix elements can be 
expressed in terms of two independent functions, $\tau(y)$ and $\zeta(y)$
\cite{IWsr,Leib}, respectively, in the limit $m_Q\to\infty$.

There are  $\Lambda_{\mathrm QCD}/m_Q$ corrections to the weak matrix elements parametrized 
by form factors at the  $m_Q\to\infty$ limit. The $\Lambda_{\mathrm QCD}/m_Q$  corrections 
to the leading term can be analyzed in a systematical way in HQET. The matrix elements 
receive contributions from higher-dimensional operators in the effective currents and in the 
effective Lagrangian. They result in a set of new Isgur-Wise functions. 

The universal functions  must be estimated
in some nonperturbative approaches. A viable approache  is
the QCD sum rules formulated in the framework of HQET. This method allows to
relate  hadronic observables to QCD parameters {\it via}
the operator product expansion (OPE) of the correlator.
A fruitful application of QCD sum rules has been the determination of
the Isgur-Wise functions parameterizing the $B\to D^{(*)}$ semileptonic
transitions up to the $\Lambda_{\mathrm QCD}/m_Q$ corrections\cite{BBBG,shifman,neubert}.

The semileptonic $B$ decays into excited charmed mesons have attracted 
attention in recent years. From the phenomenological point of view, the $B \to D^{**}$
transitions are interesting (here $D^{**}$ denotes the generic 
$L=1$ charmed state), since in principle these decay modes may  
account for a sizeable fraction of the inclusive semileptonic $B$-decay
rate. At leading order in the $1/m_Q$ expansion, the two independent universal
form factors, $\tau(y)$ and $\zeta(y)$, that parametrize the transitions
$B\to (D_1,D^*_2)\ell\bar\nu$ and $B\to (D'_0,D'_1)\ell\bar\nu$, respectively,
have been calculated with QCD sum rules\cite{c-sum,h-dai}. Moreover, perturbative corrections 
to ${\cal O}(\alpha_s)$ have been included in the QCD sum rule for $\zeta(y)$ in \cite{cnew}. 
The other approaches include different quark models\cite{godfrey,iw2,cccn,wambach,veseli,oliver,DDG} and
relativistic Bethe-Salpeter equations \cite{dai2}. A model independent 
analysis has been carried out in \cite{Leib} for the $\Lambda_{\mathrm QCD}/m_Q$ corrections.  
At the order $1/m_Q$, the corrections for  matrix elements of $B\to D^{**}$
include contributions from higher-dimensional operators in the effective currents 
and in the effective Lagrangian. For the semileptonic transitions 
$B\to D_1\ell\bar\nu$ and $B\to D^*_2\ell\bar\nu$,
the former give rise to two independent universal functions, denoted by $\tau_1(y)$
and $\tau_2(y)$\cite{Leib}. In the present work we shall focus on the first type of corrections and
investigate these two form factors, $\tau_{1,2}(y)$, with QCD sum rules in HQET. 

The remainder of this paper is organized as follows. In Sec.
\ref{sec2}  we review the formulae for the matrix elements of the weak currents including the 
structure of the $\Lambda_{\mathrm QCD}/m_Q$ corrections in the effective theory.
The QCD sum  rule analysis for the subleading  Isgur-Wise functions related to the 
corrections from matching weak currents in QCD onto those in HQET currents is presented in Sec. \ref{sec3}.   
Sec. \ref{sec4} is devoted to numerical results and  our conclusions.

%%%%%%%%%%%%%%%%%%%%%%%%%%%%%%%%%%%%%%%%%%%%%%%%%%%%%%%%%%%%%%%%%%%%%%%%%%%%% 
\section{ The heavy-quark expansion and the subleading Isgur-Wise form factors} 
\label{sec2}

The theoretical description of semileptonic decays involves the
 matrix elements of vector and axial vector currents
 ($V^\mu=\bar c\,\gamma^\mu\,b$ and $A^\mu=\bar c\,\gamma^\mu\gamma_5\,b$)
between $B$ mesons and excited $D$ mesons. For the processes $B\to D_1\ell\bar\nu$ and
$B\to D_2^*\ell\bar\nu$,
these matrix elements can be parameterized as
\begin{mathletters}\label{matrix1}
\begin{eqnarray}%\label{matrix1}
{\langle D_1(v',\epsilon)|\, V^\mu\, |B(v)\rangle}
  &=& f_{V_1}\, \epsilon^{*\mu} 
  + (f_{V_2} v^\mu + f_{V_3} v'^\mu)\, \epsilon^*\cdot v \,,  \\*
{\langle D_1(v',\epsilon)|\, A^\mu\, |B(v)\rangle  }
  &=& i\,f_A\, \varepsilon^{\mu\alpha\beta\gamma} 
  \epsilon^*_\alpha v_\beta v'_\gamma \,,  \\*
{\langle D^*_2(v',\epsilon)|\, A^\mu\, |B(v)\rangle }
  &=&k_{A_1}\, \epsilon^{*\mu\alpha} v_\alpha 
  + (k_{A_2} v^\mu + k_{A_3} v'^\mu)\,
  \epsilon^*_{\alpha\beta}\, v^\alpha v^\beta \,,  \\*
{\langle D^*_2(v',\epsilon)|\, V^\mu\, |B(v)\rangle}
  &=& i\;k_V\, \varepsilon^{\mu\alpha\beta\gamma} 
  \epsilon^*_{\alpha\sigma} v^\sigma v_\beta v'_\gamma \,.  \label{mx1-1}
\end{eqnarray}\end{mathletters}
The form factors $f_i$ and $k_i$ are functions of $y=v\cdot v'$, which 
 can be expressed by a set of Isgur-Wise
functions at each order in $\Lambda_{\rm QCD}/m_{c,b}$.
In the infinite mass limit a convenient way to evaluate hadronic matrix elements is by using the 
covariant trace formalism developed in Ref.~\cite{Falk}.
The heavy mesons can be described by spin wave-functions with
well-defined transformation properties under the Lorentz group and
heavy-quark symmetry. The spin doublet parametrized by $j_\ell^P$ can be 
represented by a $4\times 4$ Dirac-type matrix. For $j_\ell^P=1^-/2$ and 
$3^+/2$, the matrix representation are
\begin{mathletters}\label{c-rep}
\begin{eqnarray}\label{c-rep1}
&&H_v = \frac{1+\fmslash v}2\, \Big[ P_v^{*\mu} \gamma_\mu 
  - P_v\, \gamma_5 \Big] \,,\\
&&F_v^\mu = \frac{1+\fmslash v}2 \bigg\{ \! P_v^{*\mu\nu} \gamma_\nu 
  - \sqrt{\frac32}\, P_v^\nu \gamma_5 \bigg[ g^\mu_\nu - 
  \frac13 \gamma_\nu (\gamma^\mu-v^\mu) \bigg] \bigg\} \;,  \label{c-rep2}
\end{eqnarray}\end{mathletters}
where $P_v$, $P_v^{*\mu}$ and $P_v^\nu$, $P_v^{*\mu\nu}$ are annihilation
operators for members of the $j_\ell^P=1^-/2$ and $3^+/2$ doublets 
with four-velocity $v$ in HQET. The matrices $H$ and $F$ satisfy 
$\fmslash v H_v=H_v=-H_v\fmslash v$,~
$\fmslash v F_v^\mu=F_v^\mu=-F_v^\mu\fmslash v$,~ $F_v^\mu\gamma_\mu=0$, and 
$v_\mu F_v^\mu=0$.

Up to the order ${\cal O}(1/m_{c,b})$ the current $\bar c\,\Gamma\,b$
is represented in HQET by 
\begin{equation}\label{HQETcurrent}
\bar c\, \Gamma\, b = \bar h_{v'}^{(c)}\, 
  \bigg( \Gamma - \frac i{2m_c} \overleftarrow D\!\!\!\!\slash\, \Gamma  
  + \frac i{2m_b}\, \Gamma \overrightarrow D\!\!\!\!\slash
  \bigg)\, h_v^{(b)} \,,
\end{equation}
where  $h_v^{(Q)}$ is the heavy quark field in the effective theory.
Hadronic matrix elements of the leading order current between the 
states annihilated by the fields in $H_v$ and $F_{v'}^\sigma$ are
written as 
\begin{eqnarray}
\label{leading}
\bar h^{(c)}_{v'}\, \Gamma\, h^{(b)}_v = \tau\;
  {\rm Tr}\, \Big\{ v_\sigma \bar F^\sigma_{v'}\, \Gamma\, H_v \Big\} \,.   
\end{eqnarray}
Here $\tau$ is a universal Isgur-Wise function of $y$. 

At order $\Lambda_{\rm QCD}/m_{c,b}$ there are corrections originating from the matching 
of the $b\to c$ flavor changing current onto those in the effective theory.
For matrix elements between the states annihilated by the fields in 
$F_{v'}^\sigma$ and $H_v$, the new operators in Eq.~(\ref{HQETcurrent}) at the order 
$\Lambda_{\rm QCD}/m_{c,b}$ can be written as 
\begin{eqnarray}\label{curr}
\bar h^{(c)}_{v'}\, i\overleftarrow D_{\!\lambda}\, \Gamma\, h^{(b)}_v &=&
  {\rm Tr}\, \Big\{ {\cal S}^{(c)}_{\sigma\lambda}\, 
  \bar F^\sigma_{v'}\, \Gamma\, H_v \Big\} \,, \nonumber\\*
\bar h^{(c)}_{v'}\, \Gamma\, i\overrightarrow D_{\!\lambda}\, h^{(b)}_v &=&
  {\rm Tr}\, \Big\{ {\cal S}^{(b)}_{\sigma\lambda}\, 
  \bar F^\sigma_{v'}\, \Gamma\, H_v \Big\} \,.
\end{eqnarray}
The most general decomposition of the form factor is
 \begin{equation}\label{Sdef}
{\cal S}^{(Q)}_{\sigma\lambda} = v_\sigma \Big[ \tau_1^{(Q)}\, v_\lambda 
  + \tau_2^{(Q)}\, v'_\lambda + \tau_3^{(Q)}\, \gamma_\lambda \Big] 
  + \tau_4^{(Q)}\, g_{\sigma\lambda} \,.
\end{equation}
The functions $\tau_i$ depend on $y$ and have mass dimension one.  
Using the equation of motion for the heavy quark, $iv\cdot D\,h_v^{(Q)}=0$,
and translation invariance, $i\partial_\nu\,(\bar h_{v'}^{(c)}\,\Gamma\,h_v^{(b)}) =% 
(\bar\Lambda v_\nu-\bar\Lambda'v'_\nu)\,\bar h_{v'}^{(c)}\,\Gamma\,h_v^{(b)}$, 
one can obtain the constraints \cite{Leib}
\begin{eqnarray}\label{const1}
&&y\,\tau_1^{(c)} + \tau_2^{(c)} - \tau_3^{(c)} = 0 \,, \nonumber\\*
&&  \tau_1^{(b)} + y\,\tau_2^{(b)} - \tau_3^{(b)} + \tau_4^{(b)} = 0 \,.
\end{eqnarray}
and relations between $\tau_j^{(b)}$ and $\tau_j^{(c)}$
\begin{eqnarray}\label{const2}
&&\tau_1^{(c)} + \tau_1^{(b)} = \bar\Lambda\, \tau \,,\hspace{1.2cm}  
  \tau_2^{(c)} + \tau_2^{(b)}  = -\bar\Lambda'\, \tau \,, \nonumber\\
&&  \tau_3^{(c)} + \tau_3^{(b)}  = 0 \,, \hspace{1.46cm} 
  \tau_4^{(c)} + \tau_4^{(b)}  = 0 \,,
\end{eqnarray}
where $m_Q+\bar\Lambda$ and $m_Q+\bar\Lambda'$ are masses of doublets $j_\ell^P=1^-/2$ 
and $3^+/2$ in the leading order. Furthermore, combining Eqs.~
(\ref{const1}) with (\ref{const2}) yields
\begin{eqnarray}\label{const3}
\tau_3^{(c)}&=& y\, \tau_1^{(c)} + \tau_2^{(c)} \,, \nonumber\\*
\tau_4^{(c)} &=& (y-1)\, (\tau_1^{(c)} - \tau_2^{(c)})
  - (y \bar\Lambda' - \bar\Lambda)\, \tau.
\end{eqnarray}
These relations show that all  corrections to the form factors coming from
the matching of the weak currents in QCD onto those in the effective theory are
expressible in terms of $\bar\Lambda\,\tau$ and $\bar\Lambda'\,\tau$ and two of 
the coefficient functions, say, $\tau_1^{(c)}$ and $\tau_2^{(c)}$.

Applying the above relations to the $1/m_c$ correction to the vector and axial-vector currents 
.  and dropping the superscript $c$ on 
$\tau^{(c)}_{1,2}$, the corresponding matrix elements can be written as 
\begin{mathletters}\label{mc-matrix}
\begin{eqnarray}\label{mc-BD1V}
{\langle D_1(v',\epsilon)|\,\bar h^{(c)}_{v'}\, i\overleftarrow{\fmslash D}\gamma^\mu\,
 h^{(b)}_v\;|B(v)\rangle}
  &=& {1\over\sqrt{6}}\bigg\{\,\big [\;4\,(y+1)(\bar\Lambda' y- \bar\Lambda)\tau-3\,(y^2-1)
(\tau_1-\tau_2)\,\big ]\; \epsilon^{*\mu}\;\nonumber\\*
&&+\big\{\big[\,(4y-1)\tau_1+5\tau_2\big]\,v^\mu-\big[\,4(\bar\Lambda' y- \bar\Lambda)
\tau+(y+2)\tau_1\;\nonumber\\*
&&+(3y+2)\tau_2\,\big]\;v^{'\mu}\;\big\}\;\epsilon^*\cdot v\,\bigg\}\;, \\*
{\langle D_1(v',\epsilon)|\,\bar h^{(c)}_{v'}\, i\overleftarrow{\fmslash D}\gamma^\mu\gamma_5\,
 h^{(b)}_v\; |B(v)\rangle  }
  &=&  {i\over\sqrt{6}}\big[\,4(\bar\Lambda' y- \bar\Lambda)\tau-3(y-1)(\tau_1-\tau_2)\big]\,
 \varepsilon^{\mu\alpha\beta\gamma}\epsilon^*_\alpha v_\beta v'_\gamma \,, \label{mc-BD1A} \\*
{\langle D^*_2(v',\epsilon)|\,\bar h^{(c)}_{v'}\, i\overleftarrow{\fmslash D}\gamma^\mu\,
 h^{(b)}_v\; |B(v)\rangle }
  &=& i\;(\tau_1-\tau_2)\, \varepsilon^{\mu\alpha\beta\gamma} \,
  \epsilon^*_{\alpha\sigma} v^\sigma v_\beta v'_\gamma\,,\label{mc-BD2V} \\*
{\langle D^*_2(v',\epsilon)|\,\bar h^{(c)}_{v'}\, i\overleftarrow{\fmslash D}\gamma^\mu\gamma_5\,
 h^{(b)}_v\;|B(v)\rangle}
  &=&(y-1)(\tau_1-\tau_2)\, \epsilon^{*\mu\alpha}\, v_\alpha 
  + \big[\;2\tau_1 \,v^\mu +(\tau_1+\tau_2)\, v'^\mu)\,\big] \nonumber\\*&&\times\;\,
  \epsilon^*_{\alpha\beta}\, v^\alpha v^\beta \,. \label{mc-BD2A}
\end{eqnarray}
\end{mathletters}
The analogous formulae for the matrix elements of the $1/m_b$ correction to the currents can
 be written down in a similar way. 

There are also order $\Lambda_{\rm QCD}/m_{c,b}$ corrections 
originating from terms in the HQET Lagrangian of this order. They can be parametrized by
additional functions of $y$\cite{Leib}. From now on we shall confine our attention to the determination of 
subleading Isgur-Wise functions, $\tau_1(y)$ and $\tau_2(y)$, associated with the matching of
 the vector and axial vector currents in full QCD  onto those in the effective theory.

\section{Sum rules for $\tau_1$ and $\tau_2$} 
\label{sec3}

A basic element in the application of QCD sum rules to problems involving excited heavy mesons is 
to choose a set of appropriate interpolating currents  in terms of quark fields 
each of which creates (annihilate) an excited state of the heavy meson with definite quantum numbers
$j$, $P$, $j_\ell$.  The proper interpolating 
current $J_{j,P,j_{\ell}}^{\alpha_1\cdots\alpha_j}$
for the state with  arbitrary quantum number $j$, $P$, $j_{\ell}$ in HQET was
given in \cite{huang}. These currents have nice properties. 
They were proved to satisfy the following conditions 
\begin{eqnarray}
\label{decay}
\langle 0|J_{j,P,j_{\ell}}^{\alpha_1\cdots\alpha_j}(0)|j',P',j_{\ell}^{'}\rangle&=&i\,
f_{Pj_l}\delta_{jj'}
\delta_{PP'}\delta_{j_{\ell}j_{\ell}^{'}}\eta^{\alpha_1\cdots\alpha_j}\;,\\
\label{corr}
i\:\langle 0|T\left (J_{j,P,j_{\ell}}^{\alpha_1\cdots\alpha_j}(x)J_{j',P',j_{\ell}'}^{\dag
\beta_1\cdots\beta_{j'}}(0)\right )|0\rangle&=&\delta_{jj'}\delta_{PP'}\delta_{j_{\ell}j_{\ell}'}
(-1)^j\:{\cal S}\:g_t^{\alpha_1\beta_1}\cdots g_t^{\alpha_j\beta_j}\nonumber\\[2mm]&&\times\:
\int \,dt\delta(x-vt)\:\Pi_{P,j_{\ell}}(x)
\end{eqnarray}
in the $m_Q\to\infty$ limit. Where $\eta^{\alpha_1\cdots\alpha_j}$ is the 
polarization tensor for the spin $j$ state,  $v$ is the velocity of the heavy
quark, $g^{\alpha\beta}_t=g^{\alpha\beta}-v^\alpha v^\beta$ is the transverse 
metric tensor, ${\cal S}$ denotes symmetrizing the indices and
subtracting the trace terms separately in the sets $(\alpha_1\cdots\alpha_j)$
and $(\beta_1\cdots\beta_{j})$, $f_{P,j_{\ell}}$ and $\Pi_{P,j_{\ell}}$ are
a constant and a function of $x$ respectively which depend only on $P$ and $%
j_{\ell}$. Because of equations (\ref{decay}) and (\ref{corr}), the sum rules
in HQET for decay amplitudes derived from a correlator containing such currents
receive contribution only from one of the two states with the same spin-parity
$(j,P)$ but different $j_\ell$ in the $m_Q\to\infty$. Starting from the
calculations in the leading order, the decay amplitudes for finite $m_Q$ can
be calculated unambiguously order by order in the $1/m_Q$ expansion in HQET.

 Following \cite{huang} the local interpolating current 
for creating $0^-$  pseudoscalar $B$ meson is taken as
\begin{eqnarray}
\label{p-scalar}
J^{\dag\alpha}_{0,-,{1/2}}=\sqrt{\frac{1}{2}}\:\bar h_v\gamma_5q\;,
\end{eqnarray}
and the local interpolating currents for creating $1^+$ and $2^+$ 
($D_1$, $D_2^*$) mesons in the doublet ($D_1$, $D_2^*$) are taken as 
\begin{eqnarray}
\label{curr-1}
J^{\dag\alpha}_{1,+,3/2}&=&\sqrt{\frac{3}{4}}\:\bar h_v\gamma^5(-i)\left(
{\cal D}_t^{\alpha}-\frac{1}{3}\gamma_t^{\alpha}\fmslash{\cal D}_t\right)q\;,\\
\label{curr-2} 
J^{\dag\alpha_1,\alpha_2}_{2,+,3/2}&=&\sqrt{\frac{1}{2}}\:\bar h_v
\frac{(-i)}{2}\left(\gamma_t^{\alpha_1}{\cal D}_t^{\alpha_2}+
\gamma_t^{\alpha_2}{\cal D}_t^{\alpha_1}-\frac{2}{3}\;g_t^{\alpha_1\alpha_2}
\fmslash{\cal D}_t\right)q\;,
\end{eqnarray}
where ${\cal D}$ is the covariant derivative and $\gamma_{t}^\mu=\gamma^\mu-\fmslash vv^\mu$. Note
that, without the last term in the bracket in (\ref{curr-1}) the current
would couple also to the $1^+$ state in the doublet $(0^+,1^+)$ even in
the limit of infinite $m_Q$.

The QCD sum rule analysis for the subleading form factors, $\tau_1(y)$ and $\tau_2(y)$,
proceeds along the same lines as that for the leading order Isgur-Wise function $\tau(y)$. 
Here we shall  briefly outline the the procedure for determining the Isgur-Wise 
function $\tau(y)$ and refer for details to Ref. \cite{h-dai}. The idea is to study 
the analytic properties of the three-point correlators
\begin{mathletters}\label{3-point1}
\begin{eqnarray}
 i^2\int\, d^4xd^4z\,e^{i(k'\cdot x-k\cdot z)}\;\langle 0|T\left(
 J^{\nu}_{1,+,3/2}(x)\;{\cal J}^{\mu(v,v')}_{V,A}(0)\;   
 J^{\dagger}_{0,-,1/2}(z)\right)|0\rangle&=&\Xi(\omega,\omega',y)\;{\cal L}
 ^{\mu\nu}_{V,A}\;, \\
i^2\int\, d^4xd^4z\,e^{i(k'\cdot x-k\cdot z)}\;\langle 0|T\left(
 J^{\alpha\beta}_{2,+,3/2}(x)\;{\cal J}^{\mu(v,v')}_{V,A}(0)\;   
 J^{\dagger}_{0,-,1/2}(z)\right)|0\rangle&=&\Xi(\omega,\omega',y)\;{\cal L}
 ^{\mu\alpha\beta}_{V,A}\;,
\end{eqnarray}
\end{mathletters}
where ${\cal J}^{\mu(v,v')}_{V}=\bar h(v')\gamma^\mu\,h(v)$ and 
${\cal J}^{\mu(v,v')}_{A}=\bar h(v')\gamma^\mu\gamma_5\,h(v)$ are leading order vector and 
axial vector currents, respectively. 
The variables $k$, $k'$ denote residual ``off-shell" momenta which are related to the
momenta $p$ of the heavy quark in the initial 
state and $p'$ in the final state by $k=p-m_Qv$, $k'=p'-m_{Q'}v'$, 
respectively. For heavy quarks in bound states they
 are typically of order $\Lambda_{QCD}$ and remain finite in the
heavy quark limit.  ${\cal L}_{V,A}$ are  Lorentz structures  associated with
the vector and axial vector currents.

The coefficient $\Xi(\omega,\omega',y)$ in (\ref{3-point1}) is an analytic scalar 
function in the ``off-shell energies" $\omega=2v\cdot k$ and $\omega'=2v'\cdot k'$
with discontinuities for positive values of these variables. It furthermore
depends on the velocity transfer $y=v\cdot v'$, which is fixed at its physical region for
the process under consideration. By saturating the double dispersion integrals for the 
correlators in (\ref{3-point1}) with physical intermediate states in HQET, one finds the
hadronic representation of the correlator as following
 \begin{eqnarray}
\label{pole}
\Xi_{hadro}(\omega,\omega',y)={f_{-,{1\over 2}}f_{+,{3/2}}\tau(y)
\over (2\bar\Lambda_{-,{1/2}}-\omega- i\epsilon
)(2\bar\Lambda_{+,{3/2}}-\omega'- i\epsilon)}+\mbox{higher resonances} \;,
\end{eqnarray}
where $f_{P,j_\ell}$ are constants defined in (\ref{decay}),
$\bar\Lambda_{P,j_\ell}=m_{P,j_\ell}-m_Q$.
 As the result of
equation (\ref{decay}), only one state with $j^P=1^+$ or $j^P=2^+$
 contributes to (\ref{pole}), the other resonance with the same quantum
 number $j^P$ and different $j_l$ does not contribute.
 
Following the usual procedure of QCD sum rules and making double Borel transformations
in the variables $\omega$ and $\omega'$, one obtains the sum rule for $\tau$ as follows 
\begin{eqnarray}
\label{sum-tau}
\tau(y)\,f_{-,1/2}\,f_{+,3/2}\;e^{-2(\bar\Lambda_{-,1/2}/T_1
+\bar\Lambda_{+,3/2}/T_2)}=
\int_{D} d \nu d \nu^\prime \rho^{pert}(\nu,\nu^\prime,y) 
e^{-{\nu/T_1} - {\nu^\prime/T_2} }\nonumber \\* 
- {m_0^2\over 6}\;{\langle\bar qq\rangle\over T_2}%\nonumber \\* 
 -\langle \frac{\alpha_s}{\pi}GG\rangle\bigg\{
{T_1T_2[(1+2y)T_1^2+3T_2^2+(4y+2)T_1T_2]
\over 48(T_1^2+T_2^2+2yT_1T_2)^2}   \nonumber \\* 
-{(y^2-1)(T_1+T_2)T_1^3T_2^2\over 12(T_1^2+T_2^2+2yT_1T_2)^3}\bigg\}
=K_\tau(T_1,T_2,\omega_c,\omega'_c ;y)\;,\hspace{2cm}
\end{eqnarray}
where $T_1$ and $T_2$ are Borel parameters and
\begin{eqnarray}\label{rho-pert}
\rho^{pert}(\nu,\nu^\prime,y)&=&\frac{3}{2^7\pi^2}\bigg(\frac{1}{\sqrt{y^2-1}}\bigg)^3
\frac{1}{y+1}[-3\nu^2+(2y-1)(\nu^{\prime 2}+2\nu\nu')] \nonumber\\
&& \quad\mbox{}\times\Theta(\nu)\,\Theta(\nu')\,
\Theta(2y\nu\nu'-\nu^2-\nu^{'2})\;.
\end{eqnarray}
The integration domain ${ D}$ is constrained by the conditions $\nu\le \omega_c$,
$\nu^\prime  \le \omega^\prime_c$ together with the $\Theta$ functions in (\ref{rho-pert}).

Let us now turn to the derivation of the QCD sum rules for the subleading form 
factors, $\tau_1(y)$ and $\tau_2(y)$, defined in (\ref{Sdef}). To this end, we 
consider the following three-point correlation functions
\begin{mathletters}\label{3-point2}
\begin{eqnarray}
 i^2\int\, d^4xd^4z\,e^{i(k'\cdot x-k\cdot z)}\;\langle 0|T\left(
 J^{\nu}_{1,+,3/2}(x)\;\bar h^{(c)}_{v'}\, i\overleftarrow{\fmslash D}\gamma^\mu\,
 h^{(b)}_v(0)\;J^{\dagger}_{0,-,1/2}(z)\right)|0\rangle&=&\Xi_{1V}^{\mu\nu}(\omega,\omega',y)\;,
\label{3-1v} \\
 i^2\int\, d^4xd^4z\,e^{i(k'\cdot x-k\cdot z)}\;\langle 0|T\left(
 J^{\nu}_{1,+,3/2}(x)\;\bar h^{(c)}_{v'}\, i\overleftarrow{\fmslash D}\gamma^\mu\gamma_5\,
 h^{(b)}_v(0)\;J^{\dagger}_{0,-,1/2}(z)\right)|0\rangle&=&\Xi_{1A}^{\mu\nu}(\omega,\omega',y)\;,
\label{3-1a} \\
i^2\int\, d^4xd^4z\,e^{i(k'\cdot x-k\cdot z)}\;\langle 0|T\left(
 J^{\alpha\beta}_{2,+,3/2}(x)\;\bar h^{(c)}_{v'}\, i\overleftarrow{\fmslash D}\gamma^\mu\,
 h^{(b)}_v(0)\;J^{\dagger}_{0,-,1/2}(z)\right)|0\rangle&=&\Xi_{2V}^{\mu\alpha\beta}(\omega,\omega',y)\;,
\label{3-2v}\\
i^2\int\, d^4xd^4z\,e^{i(k'\cdot x-k\cdot z)}\;\langle 0|T\left(
 J^{\alpha\beta}_{2,+,3/2}(x)\;\bar h^{(c)}_{v'}\, i\overleftarrow{\fmslash D}\gamma^\mu\gamma_5\,
h^{(b)}_v(0)\;J^{\dagger}_{0,-,1/2}(z)\right)|0\rangle&=&\Xi_{2A}^{\mu\alpha\beta}(\omega,\omega',y)\;.
\label{3-2a}
\end{eqnarray}
\end{mathletters}

By saturating the double dispersion integral for the three-point functions in (\ref{3-point2}) with  hadron 
states and using (\ref{mc-matrix}) and (\ref{decay}),
one can isolate the  contributions from the double pole at 
$\omega=2\bar\Lambda_{-,{1/2}}$, $\omega'=2\bar\Lambda_{+,{3/2}}$:
\begin{mathletters}\label{Xi-pole}
\begin{eqnarray}
\Xi_{1V}^{\mu\nu}(\omega,\omega',y)&=&{f_{-,{1\over 2}}f_{+,{3/2}}\big[\,4(y\bar\Lambda_{+,{3/2}} - 
\bar\Lambda_{-,{1/2}})\tau(y){\cal L}^{\mu\nu}_{1V\tau}+\tau_1(y){\cal L}^{\mu\nu}_{1V\tau_1}
+\tau_2(y){\cal L}^{\mu\nu}_{1V\tau_2}\big]\,
\over (2\bar\Lambda_{-,{1/2}}-\omega- i\epsilon)(2\bar\Lambda_{+,{3/2}}-\omega'- i\epsilon)}+\cdots\;, \\
\Xi_{1A}^{\mu\nu}(\omega,\omega',y)&=&{f_{-,{1\over 2}}f_{+,{3/2}}\big[\,4(y\bar\Lambda_{+,{3/2}} - 
\bar\Lambda_{-,{1/2}})\tau(y)-3(y-1)(\tau_1(y)-\tau_2(y))\big]\,{\cal L}^{\mu\nu}_{1A}
\over (2\bar\Lambda_{-,{1/2}}-\omega- i\epsilon
)(2\bar\Lambda_{+,{3/2}}-\omega'- i\epsilon)}+\cdots\;,\\
\Xi_{2V}^{\mu\alpha\beta}(\omega,\omega',y)&=&{f_{-,{1\over 2}}f_{+,{3/2}}
[\tau_1(y)-\tau_2(y)]{\cal L}^{\mu\alpha\beta}_{2V}
\over (2\bar\Lambda_{-,{1/2}}-\omega- i\epsilon
)(2\bar\Lambda_{+,{3/2}}-\omega'- i\epsilon)}+\cdots\;\;,\\
\Xi_{2A}^{\mu\alpha\beta}(\omega,\omega',y)&=&{f_{-,{1\over 2}}f_{+,{3/2}}
[\tau_1(y){\cal L}^{\mu\alpha\beta}_{2A\tau_1}+\tau_2(y){\cal L}^{\mu\alpha\beta}_{2A\tau_2}]
\over (2\bar\Lambda_{-,{1/2}}-\omega- i\epsilon)(2\bar\Lambda_{+,{3/2}}-\omega'- i\epsilon)}+\cdots\;\;.
\end{eqnarray} 
\end{mathletters}
where the Lorentz structure ${\cal L}^{\mu\nu}$'s and ${\cal L}^{\mu\alpha\beta}$'s are collected in 
the Appendix.

On the other hand,it turns out that the calculation of these correlators 
with the operator product expansion (OPE) in HQET yields the following general structure
\begin{mathletters}\label{Xi-form}
\begin{eqnarray}
\Xi_{1V}^{\mu\nu}(\omega,\omega',y)&=&\Xi_{\tilde\tau}{\cal L}^{\mu\nu}_{1V\tau}+\Xi_{\tau_1}
{\cal L}^{\mu\nu}_{1V\tau_1}+\Xi_{\tau_2}{\cal L}^{\mu\nu}_{1V\tau_2}\;,\label{Xi-form1} \\
\Xi_{1A}^{\mu\nu}(\omega,\omega',y)&=&\big[\Xi_{\tilde\tau}-3(y-1)(\Xi_{\tau_1}
-\Xi_{\tau_2})\big]{\cal L}^{\mu\nu}_{1A}\;,\label{Xi-form2}\\
\Xi_{2V}^{\mu\alpha\beta}(\omega,\omega',y)&=&(\Xi_{\tau_1}-\Xi_{\tau_2}){\cal L}^{\mu\alpha\beta}_{2V}\;,
\label{Xi-form3}\\
\Xi_{2A}^{\mu\alpha\beta}(\omega,\omega',y)&=&\Xi_{\tau_1}{\cal L}^{\mu\alpha\beta}_{2A\tau_1}+
\Xi_{\tau_2}{\cal L}^{\mu\alpha\beta}_{2A\tau_2}\;.\label{Xi-form4}
\end{eqnarray}
\end{mathletters} 
where the coefficient functions $\Xi_{\tau_1}$, $\Xi_{\tau_2}$ and $\Xi_{\tilde\tau}$ are 
scalar analytic functions in the off-shell energies $\omega$ and $\omega'$.

Comparing (21) with (22) one can see that they are compatible. Therefore, we can calculate
the scalar functions $\Xi_i(\omega,\omega',y)$ with the QCD sum rules 
in HQET. From (\ref{Xi-form}) and (\ref{Xi-pole}) one can see that the sum rules for
 $\Xi_{\tau_1}$, $\Xi_{\tau_2}$ and $\Xi_{\tilde\tau}$ yield sum rules for $\tau_1(y)$, $\tau_2(y)$ and
$(y\bar\Lambda_{+,{3/2}}-\bar\Lambda_{-,{1/2}})\tau(y)$, respectively. In the 
theoretical calculation, for simplicity, the residual momentum $k$ is chosen to be parallel to 
$v$ such that $k_\mu=(k\cdot v)v_\mu$ (and similar for $k'$). The theoretical expression for 
the correlator in HQET consists of a perturbative part and contributions from vacuum condensations. 
Confining us to the leading order of perturbation and the operators with dimension $D\leq 5$ in 
OPE, the relevant Feynman diagrams are shown in Fig 1.  We shall focus, at first, on the 
coefficient functions $\Xi_{\tau_1}(\omega,\omega',y)$ and  $\Xi_{\tau_2}(\omega,\omega',y)$ to 
construct the sum rules for the subleading form factors $\tau_1(y)$ and  $\tau_2(y)$.

The spectral densities in the double dispersion integral for the perturbative diagram depicted
in Fig. 1(a) turn out to be
\begin{eqnarray}
\label{rho-tau1}
 \rho^{pert}_{\tau_1}(\nu,\nu',y)&=&\frac{3}{2^8\pi^2}\bigg(\frac{1}{\sqrt{y^2-1}}\bigg)^5
\frac{1}{y+1}\bigg[-5\nu^3+(12y-3)\nu^2\nu'-(6y^2-6y+3)\nu\nu^{'2} \nonumber\\*
&& -(2y^2-2y+1)\nu^{'3}\bigg] \Theta(\nu)\,\Theta(\nu')\,
\Theta(2y\nu\nu'-\nu^2-\nu^{'2})\;,\\
 \rho^{pert}_{\tau_2}(\nu,\nu',y)&=&\frac{3}{2^8\pi^2}\bigg(\frac{1}{\sqrt{y^2-1}}\bigg)^5
\frac{1}{y+1}\bigg[(4y-1)\nu^3-(9y^2-6y)\nu^2\nu'+(4y^3-8y^2\nonumber\\* &&+2y-1)\nu\nu^{'2}+
(2y^3-y^2+2y)\nu^{'3}\bigg]  \Theta(\nu)\,\Theta(\nu')\,
\Theta(2y\nu\nu'-\nu^2-\nu^{'2})\;.\label{rho-tau2}
\end{eqnarray}
The non-perturbative power corrections to the correlators are computed from the diagrams involving the
quark and gluon condensates in Fig. 1(b)-(c) in the Fock-Schwinger gauge $x_\mu A^\mu(x)=0$. 
We find that the only non-vanishing contribution is the
gluon  condensate. After adding the non-perturbative part and making the double Borel
transformations one obtains the sum rules for $\tau_1(y)$ and $\tau_2(y)$ as follows 
\begin{eqnarray}
\tau_1(y)\,f_{-,1/2}\,f_{+,3/2}\;e^{-2(\bar\Lambda_{-,1/2}/T_1
+\bar\Lambda_{+,3/2}/T_2)}&=&
\int_{ D} d \nu d \nu^\prime \rho^{pert}_{\tau_1}(\nu,\nu^\prime,y) 
e^{-{\nu/T_1} - {\nu^\prime/T_2} }\nonumber \\\mbox{}&& 
 -\langle \frac{\alpha_s}{\pi}GG\rangle H_{\tau_1}(T_1,T_2)\;,\label{sum-tau1}\\[2mm]
\tau_2(y)\,f_{-,1/2}\,f_{+,3/2}\;e^{-2(\bar\Lambda_{-,1/2}/T_1
+\bar\Lambda_{+,3/2}/T_2)}&=&
\int_{ D} d \nu d \nu^\prime \rho^{pert}_{\tau_2}(\nu,\nu^\prime,y) 
e^{-{\nu/T_1} - {\nu^\prime/T_2} }\nonumber \\\mbox{}&&
 -\langle \frac{\alpha_s}{\pi}GG\rangle H_{\tau_2}(T_1,T_2)\;,\label{sum-tau2}
\end{eqnarray}
where
\begin{eqnarray}
 H_{\tau_1}(T_1,T_2)&=&\frac{1}{48}T_1^3T_2\bigg[-T_1^5+4yT_1^4T_2+(8y+14)T_1^3T_2^2+(4y^2+12y+16)
T_1^2T_2^3\nonumber \\\*&&+(16y+3)T_1T_2^4+4T_2^5\bigg]/(T_1^2+2yT_1T_2+T_2^2)^4  \;,,\\
H_{\tau_2}(T_1,T_2)&=&\frac{1}{96}T_1^2T_2\bigg[(2y+1)T_1^6-(8y^2-8y)T_1^5T_2-(28y-9)T_1^4T_2^2+
-24y^2T_1^3T_2^3\nonumber \\\*&&-(8y^2+6y-11)T_1^2T_2^4+8yT_1T_2^5+3T_2^6\bigg]
/(T_1^2+2yT_1T_2+T_2^2)^4  \;.
\end{eqnarray}
The integration domain $D$ is restricted to the area in $\nu\le \omega_c$,
$\nu^\prime  \le \omega^\prime_c$. 

We have checked that the sum rules for $\tau_1(y)$ and $\tau_2(y)$ derived from
Eqs. (\ref{Xi-form1}) and (\ref{Xi-form4})
are the same and they are also consistant with
the sum rules derived for $\tau_1-\tau_2$ from Eqs. (\ref{Xi-form2}) and (\ref{Xi-form3}).

Furthermore, from the coefficient function $\Xi_{\tilde\tau}$ in (\ref{Xi-form1}) and 
(\ref{Xi-form2}) one finds the same sum rule for the
combination $(y\bar\Lambda_{+,{3/2}}-\bar\Lambda_{-,{1/2}})\tau(y)$ as follows
\begin{eqnarray}\label{sum-com}
 (y\bar\Lambda_{+,{3/2}}-\bar\Lambda_{-,{1/2}})\tau(y)\,f_{-,1/2}\,f_{+,3/2}\;e^{-2
(\bar\Lambda_{-,1/2}/T_1+\bar\Lambda_{+,3/2}/T_2)}={\hspace{2.5cm}}\nonumber\\-\frac{1}{2}\,\bigg(y
{\partial\over\partial T_2^{-1}}-{\partial\over\partial T_1^{-1}}\bigg)
K_\tau(T_1,T_2,\omega_c,\omega'_c ;y) \;,
\end{eqnarray} 
where $K_\tau$ in this equation is identical to the function $K_\tau$ in (\ref{sum-tau}).
Therefore, (\ref{sum-com}) is consistant with the sum rule (\ref{sum-tau}) in the leading order,

The above consistancy checks confirm that the correlators have the forms in 
(22) and that our method is consistant with the general analysis of Ref. \cite{Leib} described in Sec. \ref{sec2}.

In obtaining the sum rules (\ref{sum-tau1}), (\ref{sum-tau2})  and (\ref{sum-com})  
the quark-hadron duality has been assumed. In doing this, the contribution from higher hadronic 
states is  simulated by the perturbative part above some threshold energy. In the QCD 
sum rule analysis for Isgur-Wise functions for $B$ semileptonic decays into ground state $D$ 
mesons, it is argued by the authors of  \cite{neubert1,shifman,neubert} that the 
perturbative and the hadronic spectral densities can not be locally dual to each other, 
the necessary way to restore duality  is to integrate the spectral densities over the 
``off-diagonal'' variable $\nu_-=(\nu-\nu')/2$, keeping the  ``diagonal'' variable
$\nu_+=(\nu+\nu')/2$ fixed. It is in $\nu_+$ that the quark-hadron duality is assumed for
the integrated spectral densities. In doing this, for simplicity, the two Borel parameters
are taken to be equal:  $T_1 = T_2 =2T$.  We shall use the same prescription here.

The $\Theta$ functions in (\ref{rho-tau1}) and (\ref{rho-tau2}) imply that in terms of $\nu_+$
and $\nu_-$ the double discontinuities of the corrrelator are confined to the region
$-\sqrt{y^2-1}/(1+y)\;\nu_+\leq\nu_-\leq\sqrt{y^2-1}/(1+y)\;\nu_+$
and $\nu_+\geq 0$. According to our prescription an isosceles triangle
with the base $\nu_+ = \nu_c$ is retained in the integation
domain of the perturbative term in the sum rule.

In view of the asymmetry of the problem at hand with respect to the
initial and final states one may attempt to use an asymmetric triangle
in the perturbative integral. However, in that case the factor
$(y^2-1)^{3/2}$ in the denominator of (\ref{rho-tau1}) and (\ref{rho-tau2}) is not canceled
after the integration so that the Isgur-Wise function or it's
derivative will be divergent at $y=1$. Similar situation occurs for the
sum rule of the Isgur-Wise functions for the tansition between ground states
if a different domain is taken in the perturbative integal \cite{neubert}.

Putting everything together one obtains the final expressions for the QCD sum rules
\begin{eqnarray}
 \tau_1(y)\,f_{-,\frac{1}{2}}\,f_{+,\frac{3}{2}}\;e^{-(\bar\Lambda_{-,\frac{1}{2}}
+\bar\Lambda_{+,\frac{3}{2}})/T}&=&
-\frac{3}{8\pi^2}\frac{1}{(y+1)^4}\;\int_0^{\omega_c}d{\omega_+}\,
\omega_+^4\,e^{-\omega_+/T}\nonumber\\ && \quad\mbox{}+\frac{1}{3\times 2^5}  
\langle \frac{\alpha_s}{\pi}GG\rangle\frac{y+9}{(y+1)^3}\;T \;,\label{fin-tau1}\\[2mm]
\tau_2(y)\,f_{-,\frac{1}{2}}\,f_{+,\frac{3}{2}}\;e^{-(\bar\Lambda_{-,\frac{1}{2}}
+\bar\Lambda_{+,\frac{3}{2}})/T}&=&
\frac{1}{16\pi^2}\frac{5y-1}{(y+1)^4}\;\int_0^{\omega_c}d{\omega_+}\,
\omega_+^4\,e^{-\omega_+/T}\nonumber\\ && \quad\mbox{}-\frac{1}{3\times 2^5}  
\langle \frac{\alpha_s}{\pi}GG\rangle\frac{5y-3}{(y+1)^3}\;T \label{fin-tau2}\;.
\end{eqnarray} 

We end this section by noting that the QCD $O(\alpha_s)$ corrections have not been included 
in the sum rule calculations.  However,  the Isgur-Wise function obtained from the QCD sum rule 
actually is the ratio of the three-point correlator to the two-point correlator.  While 
both of these correlators subject to large perturbative QCD corrections, it is expected that 
their ratio is not much affected by these corrections because of cancelation.   
This has been  proved to be true in the analysis for $B$ semileptonic decay to ground state 
heavy mesons \cite{neubert}.

\section{Numerical analysis and conclusion} 
\label{sec4}

In order to obtain the values for $\tau_1(y)$ and $\tau_2(y)$ from Eqs. 
(\ref{fin-tau1}) and (\ref{fin-tau2}) in the numerical evaluation we need to use the hadronic parameters 
$\bar\Lambda$'s and $f$'s of the corresponding interpolating currents  
as input. The QCD sum rules derived from the two-point correlator has been applied
to determine $\bar\Lambda$ and $f$. $\bar\Lambda_{-,1/2}$ and 
$f_{-,1/2}$ can be obtained from the results in \cite{neubert} as 
$\bar\Lambda_{-,1/2}=0.5$ GeV and $f_{-,1/2}\simeq 0.24$ GeV$^{3/2}$ at the
order $\alpha_s=0$. Notice that the coupling constant $f_{-,1/2}$ defined in
the present work is a factor $1/\sqrt 2$ smaller than that defined in \cite
{neubert}. Determination of $\bar\Lambda_{+,3/2}$ and $f_{+,3/2}$ 
 by QCD sum rules gave the results:
$\bar\Lambda_{+,3/2}=0.95$ GeV and $f_{+,3/2}=0.19$ GeV$^{5/2}$ at the order $\alpha_s=0$ \cite{huang,hhh}.
For the QCD parameters entering the theoretical expressions, we take
the standard values: $\langle\alpha_s GG\rangle= (0.04)~\mbox{GeV}^4$.

Imposing usual criterium that both higher-order power corrections and the contribution of the 
continuum should not be very large, we find an acceptable stability window for the threshold 
parameter in the range 
$\omega_c=2.0-2.6$ GeV, in which the results do not appreciably depend on the Borel
parameter in the range $T=0.7-1.1$ GeV. The range of Borel parameter here overlaps with that
of the corresponding two-point sum rules \cite{neubert,huang}.

The  values of the form factors $\tau_1(y)$ and
$\tau_2(y)$ at zeor recoil as  functions of the Borel parameter are shown in Fig. 2(a) and 3(a), for three 
different values of the continuum
threshold $\omega_c$.  The numerical results for $\tau_1(y)$ and $\tau_2(y)$
are shown in Fig. 2(b) and 3(b), where the curves refer to three different
values of $\omega_c$ and $T$ is fixed at $T=0.9$ GeV.

The numerical analysis shows that $\tau_1(y)$ and $\tau_2(y)$ are slowly varying functions
in the allowed kinematic range for $B\to D_1\ell\bar\nu$ and $B\to D_2^*\ell\bar\nu$ decays. 
The resulting curves for $\tau_1(y)$ and $\tau_2(y)$ may be well parameterized by
 the linear approximations
\begin{eqnarray}\label{simu}
\tau_1(y)&=&\tau_1(1)\;(1-\rho_{\tau_1}^2(y-1))\;, \hspace{0.3cm}
\tau_1(1)=-0.4\pm 0.1\;, \hspace{0.4cm} \rho_{\tau_1}^2=1.4\pm 0.2 \;,\\
\tau_2(y)&=&\tau_2(1)\;(1-\rho_{\tau_2}^2(y-1))\;, \hspace{0.3cm}
\tau_2(1)=0.28\pm 0.05\;, \hspace{0.3cm} \rho_{\tau_2}^2=0.5\pm 0.1\;.
\end{eqnarray}
The errors here reflect only the uncertainty due to $\omega_c$ and $T$. They
do not contain other errors in the QCD sum rule approach.

In conclusion, we have presented a QCD sum rule analysis of the subleading Isgur-Wise
functions $\tau_1(y)$ and $\tau_2(y)$, appearing in the heavy quark expansion of the transition
 matrix elements between heavy mesons  due to matching of
the weak currents in QCD onto those in the effective theory at the order $1/m_Q$. Our approach
is in accordance with the general relations obtained from analysis based on
HQET in \cite{Leib}.

%%%%%%%%%%%%%%%%%%%%%%%%%%%%%%%%%%%%%%%%%%%%%%%%%%%%%%%%%%%%%%%%%%%%%%%%%%%%%%%%%%%%%%%%%%%%%%%%%%

\acknowledgments This work was supported in part by the National Natural Science 
Foundation of China.
%\vspace{0.5cm}
%%%%%%%%%%%%%%%%%%%%%%%%%%%%%%%%%%%%%%%%%%%%%%%%%%%%%%%%%%%%%%%%%%%%%%%%%%%%%%%%%%%%%%%%%%%%%%%%%%
\newpage
\appendix
\section*{}
We list here the  lorentz structures used in the paper.
\begin{eqnarray}
{\cal L}^{\mu\nu}_{1V\tau}&=&\frac{1}{\sqrt{6}}\big[-(y+1)g_t^{\mu\nu}+v^{'\mu}v_t^\nu\big]\;,\\
{\cal L}^{\mu\nu}_{1V\tau_1}&=&\frac{1}{\sqrt{6}}\bigg\{3(y^2-1)g_t^{\mu\nu}
-[(4y-1)v^\mu-(y+2)v^{'\mu}]v_t^\nu\bigg\}\;,\\
{\cal L}^{\mu\nu}_{1V\tau_2}&=&\frac{1}{\sqrt{6}}\bigg\{-3(y^2-1)g_t^{\mu\nu}
-[5v^\mu-(3y+2)v^{'\nu}]v_t^\nu\bigg\}\;,\\
{\cal L}^{\mu\nu}_{1A}&=&\frac{-i}{\sqrt{6}}\varepsilon^{\mu\nu\alpha\beta}v_\alpha v'_\beta\;,\\
{\cal L}^{\mu\alpha\beta}_{2V}&=&\frac{i}{2}(
 \varepsilon^{\mu\alpha\rho\sigma}v^t_\beta+\varepsilon^{\mu\beta\rho\sigma}v^t_\alpha)
v_\rho v'_\sigma \,,\\
{\cal L}^{\mu\alpha\beta}_{2A\tau_1}&=&(y-1)\bigg\{\frac{1}{2}(g_t^{\alpha\mu}v_t^\beta
+g_t^{\beta\mu}v_t^\alpha)-\frac{1}{3}g_t^{\alpha\beta}v_t^\mu\bigg\}+(2v^\mu+v^{'\mu})
\bigg\{v_t^\alpha v_t^\beta-\frac{1}{3}(1-y^2)g_t^{\alpha\beta}\bigg\}\;,\\
{\cal L}^{\mu\alpha\beta}_{2A\tau_2}&=&-(y-1)\bigg\{\frac{1}{2}(g_t^{\alpha\mu}v_t^\beta
+g_t^{\beta\mu}v_t^\alpha)-\frac{1}{3}g_t^{\alpha\beta}v_t^\mu\bigg\}+v^{'\mu}
\bigg\{v_t^\alpha v_t^\beta-\frac{1}{3}(1-y^2)g_t^{\alpha\beta}\bigg\}\;,
\end{eqnarray} 
where $g_t^{\alpha\beta}=g^{\alpha\beta}-v^{'\alpha} v^{'\beta}$ and 
$v_t^\alpha=v^\alpha-yv'^\alpha$. It is easy to see that these Lorentz structures
satisfy
\begin{eqnarray}
v'_\nu{\cal L}^{\mu\nu}_{1V\tau}=v'_\nu{\cal L}^{\mu\nu}_{1V\tau_1}=v'_\nu
{\cal L}^{\mu\nu}_{1V\tau_2}=v'_\nu{\cal L}^{\mu\nu}_{1A}=0\;\,\\
g_{\alpha\beta}{\cal L}^{\mu\alpha\beta}_{2V}=g_{\alpha\beta}{\cal L}^{\mu\alpha\beta}
_{2A\tau_1}=g_{\alpha\beta}{\cal L}^{\mu\alpha\beta}_{2A\tau_2}=0\;\,\\
v'_\alpha{\cal L}^{\mu\alpha\beta}_{2V}=v'_\alpha{\cal L}^{\mu\alpha\beta}_{2A\tau_1}=
v'_\alpha{\cal L}^{\mu\alpha\beta}_{2A\tau_2}=0\;,\\
v'_\beta{\cal L}^{\mu\alpha\beta}_{2V}=v'_\beta{\cal L}^{\mu\alpha\beta}_{2A\tau_1}=
v'_\beta{\cal L}^{\mu\alpha\beta}_{2A\tau_2}=0\;.
%v'_\alpha{\cal L}^{\mu\alpha\beta}_{2V}=v'_\beta{\cal L}^{\mu\alpha\beta}_{2V}=
%v'_\alpha{\cal L}^{\mu\alpha\beta}_{2A\tau_1}=v'_\beta{\cal L}^{\mu\alpha\beta}_{2A\tau_1}=
%v'_\alpha{\cal L}^{\mu\alpha\beta}_{2A\tau_2}=v'_\beta{\cal L}^{\mu\alpha\beta}_{2A\tau_2}=0\;.
\end{eqnarray} 
The appearence of the Lorentz structures satisfying these relations is the result of the 
following equations
\begin{eqnarray}
&&v'_\nu\;J^\nu_{1,+,3/2}=v'_\alpha\;J^{\alpha\beta}_{2,+,3/2}=0\;,\\
&&g_{\alpha\beta}\;J^{\alpha\beta}_{2,+,3/2}=0\;,\hspace{0.4cm}
J^{\alpha\beta}_{2,+,3/2}=J^{\beta\alpha}_{2,+,3/2}\;,
\end{eqnarray} 
satisfied by the interpolating currents in the correlators.

%%%%%%%%%%%%%%%%%%%%%%%%%%%%%%%%%%%%%%%%%%%%%%%%%%%%%%%%%%%%%%%%%%%%%%%%%%%%%%%%%%%%%%%%%%%%%%%%%%

\newpage
{\bf Figure Captions}
\vspace{2ex}
\begin{center}
\begin{minipage}{120mm}
{\sf Fig. 1.} \small{Feynman diagrams contributing to the sum rules for the 
                    Isgur-Wise form factors in the coordinate gauge. The gray
                    square corresponds to the insertion of the $1/m_c$ current.  }
\end{minipage}\end{center}

\begin{center}
\begin{minipage}{120mm}
%\begin{minipage}{120mm}
{\sf Fig. 2.} \small{ Numerical evaluation for the sum rule (\ref{fin-tau1}): (a)
dependence of $\tau_1(1)$ on the Borel parameter $T$ for defferent values of the 
continuum threshold $\omega_c$; (b) Results for the  Isgur-Wise form factor $\tau_1(y)$
with $T=0.9$ GeV.}
\end{minipage}\end{center}
%\end{document}

\begin{center}
\begin{minipage}{120mm}
%\begin{minipage}{120mm}
{\sf Fig. 3.} \small{Numerical evaluation for the sum rule (\ref{fin-tau2}): (a)
dependence of $\tau_2(1)$ on the Borel parameter $T$ for defferent values of the 
continuum threshold $\omega_c$; (b) Results for the  Isgur-Wise form factor $\tau_2(y)$
with $T=0.9$ GeV.}
\end{minipage}\end{center}
%\end{document}
\vspace{0.8cm}
  
%%%%%%%%%%%%%%%%Fig. 1%%%%%%%%%%%%%%%%%%55 
\begin{figure}[htbp]   % produce figure here 
\begin{center}
\setlength{\unitlength}{1truecm} 
\begin{picture}(6.8,6.8)%(<right,>top) 
\put(-8.0,-24)
{\includegraphics{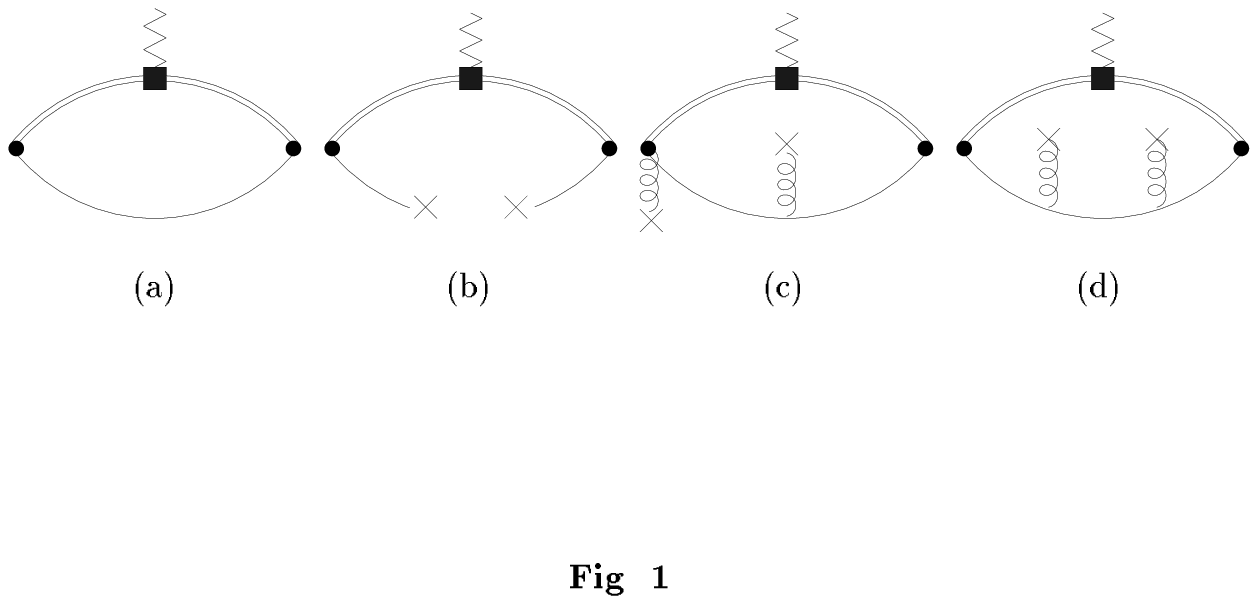}} 
\end{picture} 
\end{center} 
\vskip 2.0cm 
%\fcaption{xx}
\protect\label{Fig.1}
\end{figure}
\newpage
%%%%%%%%%%%%%%%%Fig. 2a%%%%%%%%%%%%%%%%%%%
\begin{figure}[htbp]   % produce figure here 
\begin{center}
\setlength{\unitlength}{1truecm} 
\begin{picture}(6.8,6.8)%(<right,>top) 
\put(-8.0,-14)
{\includegraphics{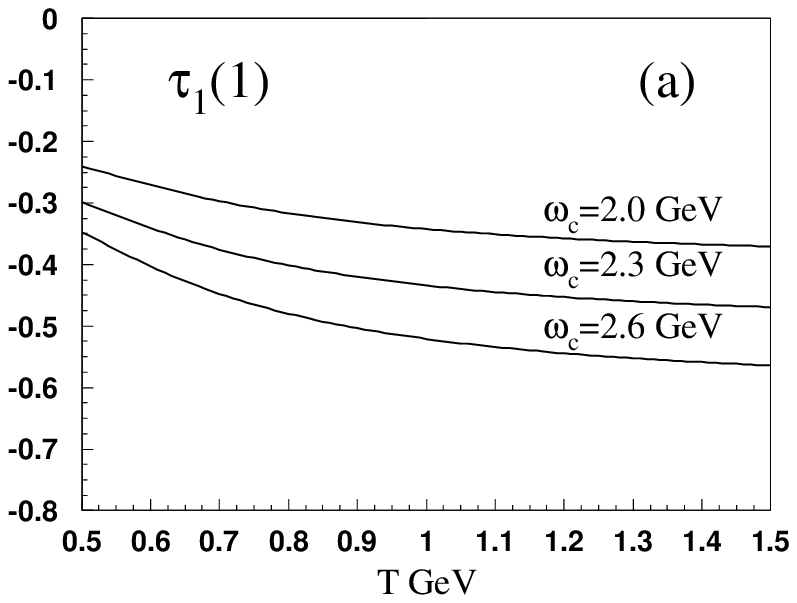}} 
\end{picture} 
\end{center} 
\vskip 2.0cm 
%\fcaption{xx}
\protect\label{Fig.2a}
\end{figure}
%%%%%%%%%%%%%%%%Fig. 2b%%%%%%%%%%%%%%%%%%%
\begin{figure}[htbp]   % produce figure here 
\begin{center}
\setlength{\unitlength}{1truecm} 
\begin{picture}(6.8,6.8)%(<right,>top) 
\put(-8.0,-9.9)
{\includegraphics{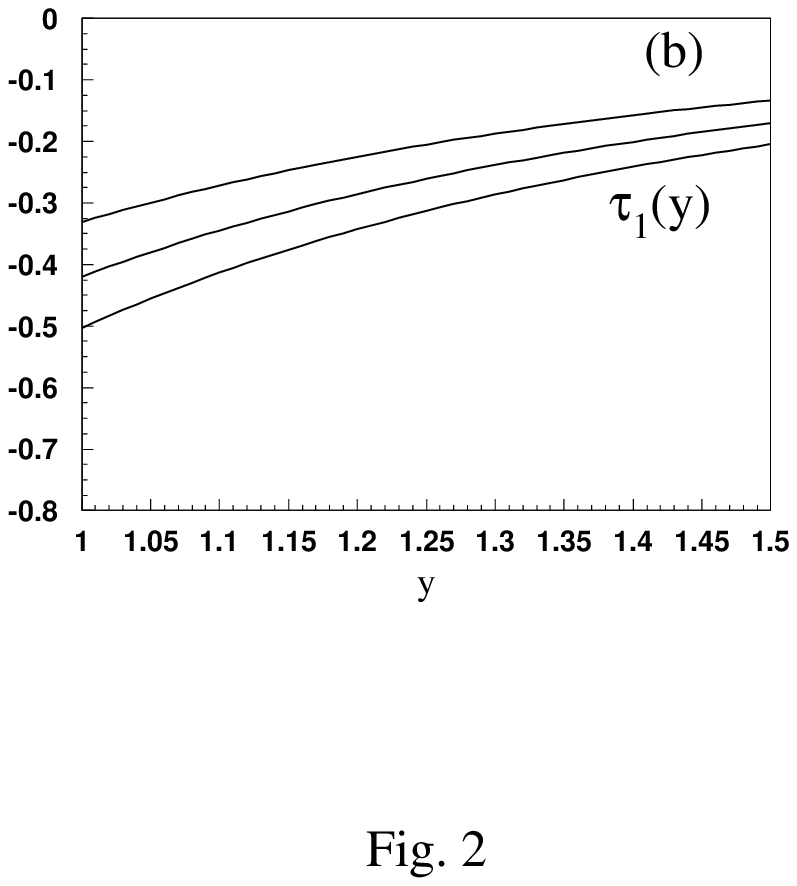}} 
\end{picture} 
\end{center} 
\vskip 2.0cm 
%\fcaption{xx}
\protect\label{Fig.2b}
\end{figure}

%%%%%%%%%%%%%%%%Fig. 3a%%%%%%%%%%%%%%%%%%%
\begin{figure}[htbp]   % produce figure here 
\begin{center}
\setlength{\unitlength}{1truecm} 
\begin{picture}(6.8,6.8)%(<right,>top) 
\put(-8.0,-14)
{\includegraphics{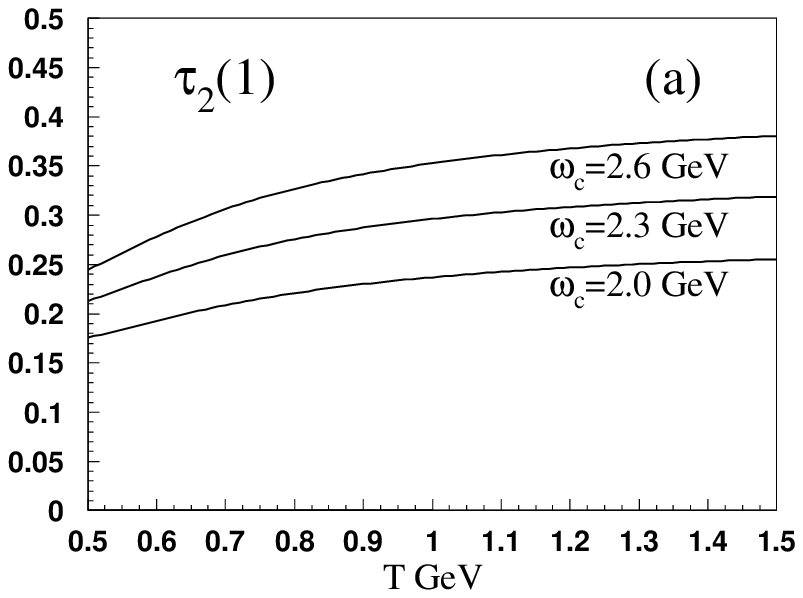}} 
\end{picture} 
\end{center} 
\vskip 2.0cm 
%\fcaption{xx}
\protect\label{Fig.3a}
\end{figure}
%%%%%%%%%%%%%%%%Fig. 3b%%%%%%%%%%%%%%%%%%%
\begin{figure}[htbp]   % produce figure here 
\begin{center}
\setlength{\unitlength}{1truecm} 
\begin{picture}(6.8,6.8)%(<right,>top) 
\put(-8.0,-9.9)
{\includegraphics{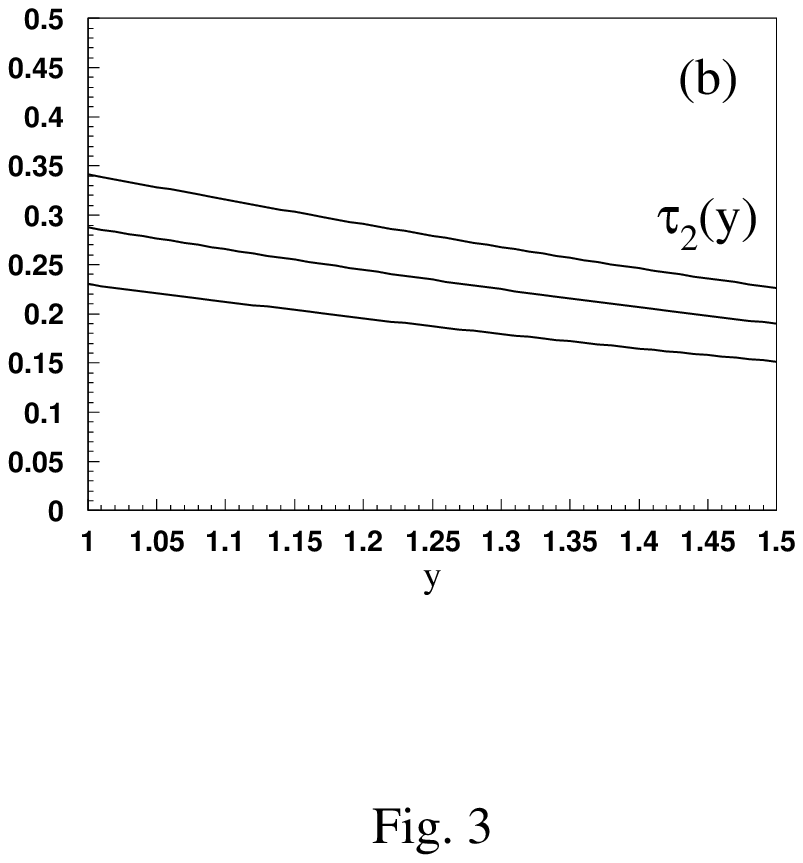}} 
\end{picture} 
\end{center} 
\vskip 2.0cm 
%\fcaption{xx}
\protect\label{Fig.3b}
\end{figure}
 
\end{document}